\documentclass[prd,twocolumn,showpacs,amsmath,amssymb]{revtex4}
\usepackage{graphicx}
\usepackage{dcolumn}
\usepackage{bm}
\def\br{\begin{eqnarray}}
\def\er{\end{eqnarray}}
\def\be{\begin{equation}}
\def\ee{\end{equation}}

\def\({\left(}
\def\){\right)}

\begin{document}

\title{Light composite Higgs from an effective action for Technicolor}
                 
\author{A.~Doff}
\email{agomes@utfpr.edu.br}
\affiliation{Universidade Tecnol\'ogica Federal do Paran\'a - UTFPR - COMAT\\
Via do Conhecimento Km 01, 85503-390, Pato Branco - PR, Brazil}

\author{A.~A.~Natale}
\email{natale@ift.unesp.br}
\affiliation{Instituto de F\'{\i}sica Te\'orica, UNESP \\
Rua Pamplona, 145,
01405-900, S\~ao Paulo - SP,
Brazil}

\author{P.~S.~Rodrigues da Silva}
\email{psilva@fisica.ufpb.br}
\affiliation{Departamento de F\'{\i}sica, Universidade Federal da
Para\'{\i}ba, \\ 
Caixa Postal 5008, 58051-970, Jo\~ao Pessoa - PB, Brazil}

\date{\today}

\begin{abstract}
We compute an effective action for a composite Higgs boson formed
by new fermions belonging to a general technicolor non-Abelian gauge theory, using a quite general
expression for the fermionic self-energy that depends on a certain parameter ($\alpha$),
that defines the technicolor theory from the extreme walking behavior up to the one with a
standard operator product expansion behavior. We discuss the values of the
trilinear and quadrilinear scalar couplings. Our calculation spans all the possible
physical possibilities for mass and couplings of the composite system. In the case of extreme walking technicolor
theories we verify that it is possible to have a composite Higgs boson with a mass as light as the present experimental limit,
contrary to the usual expectation of a heavy mass for the composite Higgs boson. In this case we
obtain an upper limit for the Higgs boson mass, ($M_H \leq {\cal O} (700)$~GeV for $SU(2)_{TC}$), and
the experimental data on the Higgs boson mass constrain $SU(N)_{TC}$ technicolor gauge groups to be smaller 
than $SU(10)_{TC}$.
\end{abstract}


\maketitle

\section{Introduction}

\noindent In the standard model of elementary particles the fermion and gauge boson
masses are generated due to the interaction of these particles with elementary
Higgs scalar bosons. Despite its success there are some points in the model
as, for instance, the enormous range of masses between the lightest and heaviest
fermions and other peculiarities that could be better explained at a deeper level. The nature of the Higgs boson is one of the most important problems in particle physics, and there are many questions that may be answered in the near future by the LHC experiments, such as: Is the Higgs boson, if it exists at all, elementary or composite? What are
the symmetries behind the Higgs mechanism?

The possibility that the Higgs boson is a composite state instead
of an elementary one is more akin to the phenomenon of spontaneous symmetry breaking 
that originated from the effective Ginzburg-Landau Lagrangian, which can be derived 
from the microscopic BCS theory of superconductivity describing the electron-hole 
interaction (or the composite state in our case). This dynamical origin of the
spontaneous symmetry breaking has been discussed with the use of many
models, the most popular one being the technicolor (TC) model~\cite{lane}.
Unfortunately we do not know the dynamics that form the scalar bound state, which 
should play the role of the Higgs boson in the standard model symmetry breaking, 
and no phenomenologically satisfactory model along this line has been derived 
up to now.

Most of the models for the spontaneous symmetry breaking of the standard model
based on the composite Higgs boson system depends on specific assumptions about 
the theory particle content and consequently on the dynamics
responsible for the bound state formation~\cite{hs}, 
and one of the questions that we address in
this work is how can we make predictions about the effective Higgs Lagrangian without
assuming specific models or dynamics? In principle, new fermions are bounded by
a new interaction stronger than QCD and originate a composite scalar state 
whose wave function is a solution of the Bethe-Salpeter equation. In non-Abelian
gauge theories this wave function (or the Bethe-Salpeter kernel $\Phi_{BS} (p,q)$)
is related to the self-energy of
the new fermions~\cite{delbourgo}
\be
\Sigma (p^2) =  \Phi_{BS} (p,q)|_{q \rightarrow 0} \, ,
\label{eq0}
\ee 
and here we shall assume for this self-energy 
($\Sigma (p,\alpha)$) a very 
general expression that interpolates between all possible scalar wave functions
(or all possible non-Abelian gauge group dynamics) as we vary a specific parameter
($\alpha$) present in this function~\cite{doff1}. When this parameter goes
to $1$ we obtain a fermionic self-energy that behaves as $\Sigma (p^2) \propto \Lambda^3 / p^2$,
which is the usual operator product expansion (OPE) behavior for a gauge theory
that develops a dynamical mass scale $\Lambda$~\cite{lane2}. When $\alpha \rightarrow 0$ the self-energy
is the one that appears in the extreme walking technicolor theories~\cite{walk}.
Using this self-energy ansatz we can study
several properties of the composite Higgs boson in a model
independent way~\cite{doff2}, as we choose the free parameter ($\alpha$) which defines the theory to
be considered.

Observe that Eq.(1) shows that there is formal 
relation between the fermion
self-energy ($\Sigma (p^2)$) and the scalar boson
wave function. In principle this means that if we know $\Sigma (p^2)$ we
know all the properties of the
scalar boson. However we must keep in mind that the calculation of the
effective action is not performed
with an exact expression for the self-energy (or scalar wave function),
but with a simple approximation of this
function which obeys the leading order solution of  the Schwinger-Dyson
equation (SDE) for the fermion propagator.
By leading order SDE in the case of non-Abelian gauge theories we
understand that the SDE are solved using
as input the bare gauge boson propagator and solely the effect of the
running coupling in the vertex function.
This approximation is usually assumed as reasonable and has already 
been tested at higher order for walking technicolor theories~\cite{ladder}.
On the other hand we also point out that the effective action is a
dressed loop expansion, which is able to capture the nonlinearities
of the dynamical symmetry breaking under a controllable approximation,
as shown by Cornwall and Shellard in Ref.~\cite{cs}, but
if we neglect the next order of the loop expansion and consider the
leading order and simple approximation for the fermionic self-energy
we will surely have an uncertainty in the boson masses and couplings
that we quote.

With the general self-energy (or composite state wave function) we can compute an
effective action ($\Omega$) for composite operators~\cite{cjt} 
of the effective Higgs system, which
is a type of calculation already performed for several specific models (see, for instance,
Ref.~\cite{peskin,natale}). However the effective potential has not been computed
up to now with the general self-energy ansatz that we referred to above. Moreover, the
effective potential  by itself does not give the
full information about the composite Higgs system. The effective action
contains a kinetic term, which, as demonstrated by Cornwall and Shellard~\cite{cs},
has the form
\be 
\Omega_K = \frac{1}{2} \int d^4x \, \frac{[\partial_\mu \phi (x)]^2}{\kappa} \, , 
\label{eq1}
\ee 
where $\phi$ is related to the composite wave-function and to obtain a conventional
kinetic term we define
\be
\Phi (x) = Z^{-1/2} \phi (x) \, , 
\label{eq2}
\ee
where $\Phi$ plays the role of the physical field and $Z = \kappa$ acts as a 
renormalization constant. The constant $Z$ is important to set the right
scale in our ``Ginzburg-Landau" effective Lagrangian, actually it will be
fundamental to the results in order to provide the right values of the composite
scalar boson mass and self-coupling constants. This effective Lagrangian
will be useful to set limits on the composite Higgs boson system in a quite
general way, and it will be given by $\Omega$ which is composed by
the kinetic term $\Omega_K$ and the effective potential part $\Omega_V$.
Another point that, as far as we know, has not been extensively discussed in
the literature and we discuss here are the different contributions to the effective
potential that come from the new fermions that form the scalar composite
state, and the ones that come from ordinary fermions. Both
contributions are responsible, as we shall see, for determining the value of the
composite Higgs boson mass which, as our result indicates, can be as light as few hundreds GeV, corroborating the results of  Ref.~\cite{LCH}.

This paper is organized as follows: In Sec. II we discuss the effective
potential for composite operators and how the
kinetic term of the effective action is generated through the use of
the general self-energy ansatz. 
Sec. III contains the actual calculation of the effective
action. In Sec. IV we gather our results and compute the Higgs boson masses, and in
Sec. V we draw our conclusions.

\section{Effective action and fermion self-energy}

The effective action for composite operators~\cite{cjt} ($\bar{\Gamma}$), is a function of the Green
functions $G_i$, and is stationary with respect to variations of $G_i$:
\be
\frac{\delta \bar{\Gamma}}{\delta G_i} =0 \, .
\label{eq3}
\ee 
The effective potential is defined by
\be
V(G_i) \int d^4 x \, = \, - \left. \bar{\Gamma} (G_i)\right|_{translation \, invariant} \, .
\label{eq4}
\ee
In terms of the complete fermion ($S$) and gauge boson ($D$) propagators,  $V(G_i)$ can be written as
\br
V(S,D) &=& - i \, \int \, \frac{d^4p}{(2\pi)^4} \, Tr \left( \ln S_0^{-1} S - S_0^{-1} S +1 \right) \nonumber \\
&&+ V_2 (S,D) \, ,
\label{eq5}
\er
where $S_0$ (and $D_0$) stands for the bare fermion (gauge boson) propagator.

\begin{figure}[h]
\centering
\includegraphics[width=0.7\columnwidth]{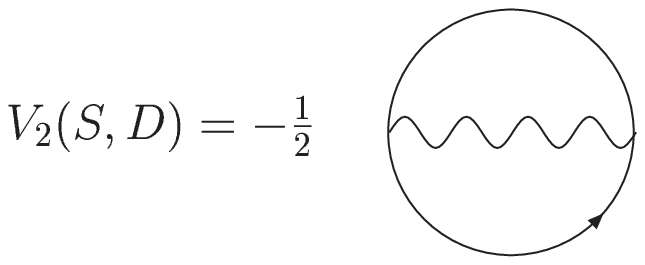}
\caption[dummy0]{Two-particle irreducible contribution to the vacuum energy.} \label{fig1}
\end{figure}

$V_2 (S,D)$ is the sum of all two-particle irreducible vacuum diagrams. The only 
contribution that we shall consider to $V_2(S,D)$ is the one depicted in Fig.(\ref{fig1}),
and the equation
\be
\frac{\delta V}{\delta S} = 0\, ,
\label{eq6}
\ee
gives the SDE for the fermion propagator. We are not considering
contributions to the potential due to gauge and ghosts loops, because we are interested
only in the fermionic bilinear condensation in the scalar channel, keeping in mind
that we should consider a non-Abelian gauge theory, stronger than QCD, whose fermions form the composite Higgs boson. Of course, we are also not considering the possibility of
gauge boson mass generation in this non-Abelian theory, as may happen in QCD~\cite{corn2}, that could imply only in a change of the potential value at the minimum, 
but not in the symmetry breaking pattern of the effective Higgs theory.   

We can represent $V_2 (S,D)$ analytically in the Hartree-Fock approximation by
\be
i V_2 (S,D) = -\frac{1}{2} \, Tr (\Gamma S\Gamma S D) \, , 
\label{eq7}
\ee
where $\Gamma$ is the fermion proper vertex. In Eq.(\ref{eq7}) we have not written the gauge
and Lorentz indices, as well as the momentum integrals.

The physically meaningful quantity that we must compute is the vacuum energy density
given by
\be
\Omega_V = V(S,D) - V(S_0, D_0) \, ,
\label{eq8}
\ee
where we are subtracting the symmetric part of the potential from the potential
that admits condensation in the scalar channel, 
that is denoted by $V(S_0, D_0)$ and is a function of the
perturbative propagators ($S_0$ and $D_0$), where the complete propagator $S$ is related to the
free propagator by
\be
S^{-1} = S_0^{-1} - \Sigma \, ,
\label{eq9}
\ee
where $S_0 = i / \not\!p$.
 
The vacuum energy density, if we remove all indices and integrations, can be 
written as~\cite{cjt,natale}
\br
\Omega_V &=& - i Tr (\ln S_0^{-1}S - S_0^{-1}S +1) + i Tr \Sigma (S-S_0) \nonumber \\
&+& \frac{1}{2} i Tr (\Gamma S \Gamma S - \Gamma S_0 \Gamma S_0)D  \, .
\label{eq10}
\er
Using Eq.(\ref{eq9}) and assuming $\Sigma S_0$ small, it is possible to expand $\Omega_V$ in powers of $\Sigma$, 
that gives~\cite{cjt,natale}
\br
\Omega_V &=&  i Tr \ln (1- \Sigma S_0) + \frac{1}{2} i Tr \Sigma S_0 \Sigma S_0 \nonumber \\
&+& \frac{1}{2} i Tr S_0\Sigma S_0 \Sigma S_0\Gamma S_0 \Sigma S_0 \Sigma S_0 \Gamma D_0  \, .
\label{eq11}
\er
In Eq.(\ref{eq11}) we have kept terms only up to the $\Sigma^4$ term that comes from the two-loop contribution.
Note that expanding the logarithmic term the $\Sigma^2$ contribution is absent, which is a consequence
of the fact that $\Sigma$ obeys the linear homogeneous SDE for the fermion propagator~\cite{cjt,cn}. 

We parametrize the self-energy $\Sigma$ as~\cite{doff1}
\be 
\Sigma (p^2) \sim \Lambda \left( \frac{ \Lambda^2 }{p^2}\right)^{\alpha}\left[1 + b g^2 \ln\left(p^2/ \Lambda^2 \right) \right]^{-\gamma\cos (\alpha \pi)}  \, .
\label{eq12}
\ee	
In the above expression $\Lambda$ is the characteristic scale of mass generation of the theory forming the
composite Higgs boson, which hereafter will be identified with the TC scale, $\Lambda_{TC}$. $b$ is the coefficient of the $g^3$ term in the renormalization group $\beta$ function,
$\gamma= 3c/16\pi^2 b$, and  $c$ is the quadratic Casimir operator given by 
$$ 
c = \frac{1}{2}\left[C_{2}(R_{1}) +  C_{2}(R_{1}) - C_{2}(R_{3})\right]\,,
$$ 
where $C_{2}(R_{i})$,  are the Casimir operators for fermions in the representations  $R_{1}$ and 
$R_{2}$ that form a composite boson in the representation $R_{3}$. 
The only restriction on this ansatz is $\gamma > 1/2$~\cite{lane2}, and if we consider the formal
equivalence between the solution of the SDE with the Bethe-Salpeter one for
scalar bound states, the above restriction indicates a condition on the composite wave-function 
normalization.

The ansatz in Eq.(\ref{eq12}), proposed in Ref.~\cite{doff1}, interpolates between  the standard OPE result for the technifermion self-energy, which is  obtained when $\alpha \rightarrow 1$, and the extreme walking technicolor solution obtained when $\alpha \rightarrow 0$~\cite{walk}, i.e., this is the case where the symmetry breaking is dominated by higher order interactions that are relevant at or above the TC scale, leading naturally to a very hard dynamics~\cite{soni,soni2}.
As two of us have pointed out in Ref.~\cite{doff2} only such kind of solution  is naturally  capable of generating a large mass to the third fermionic generation, which has a mass limit almost saturated by the top quark mass.
This variation of the ansatz with $\alpha$ is what makes our calculation a general one; it covers all
possible solutions of the SDE (or Bethe-Salpeter equation) for fermions forming
the composite boson.

We can now determine a complete effective theory (or the Ginzburg-Landau Lagrangian), including the
kinetic term of the effective action. To start with, let us suppose that the real vacuum leads to 
fermion condensation and denote the true ground state by $|\Omega \rangle$. Taking into account the
structure of the real vacuum, the fermion propagators are described by a fermion bilinear which is
not translationally invariant
\be
S(x,y)_{\eta\xi} = - i \langle \Omega | T [\chi_{\eta}(x+\frac{1}{2} y)
\psi_{\xi}(x-\frac{1}{2} y) ] |\Omega \rangle \, .  
\label{eq13}
\ee
The Fourier transform of Eq.(\ref{eq13}) can be written as
\be
S(p,k) = S_0 (p,k) + \Sigma (p,k) \, ,
\label{eq14}
\ee
where $S_0 (p,k)$ is the bare propagator (which is translationally invariant) given by
\be
S_0 (p,k) = (2\pi)^4 \delta^4 (p-k)/ \not\!k  \, ,
\label{eq15}
\ee
and $\Sigma (p,k)$ is a gap equation, which can be separated in its regular part ($\Sigma_R$ -- one
that does not represent symmetry breaking) and a singular part that breaks the symmetry ($\Sigma_S (p,k)$)
\be
\Sigma(p,k) = (2\pi)^4 \Sigma_R (k) \delta^4(p-k) + \Sigma_S(p,k) \, .
\label{eq16}
\ee
Our ansatz for $\Sigma (p^2)$ that appears in Eq.(\ref{eq12}) is nothing else than the
linearized solution of $\Sigma (p,k)$.

\begin{figure}
\centering
\includegraphics[width=0.99\columnwidth]{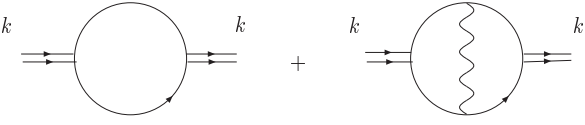}
\caption[dummy0]{Diagrams contributing to the kinetic term in the effective Lagrangian.} \label{fig2}
\end{figure}

If we suppose that the expectation value of the fermion bilinear has the following operator
expansion~\cite{cs}
\be
\langle \Omega | T [\chi (x+\frac{1}{2} y)
\psi (x-\frac{1}{2} y) ] |\Omega \rangle \, {}^{\,\, \sim}_{y \rightarrow 0}\,  C(y) \phi (x)  ,
\label{eq17}
\ee
where $C(y)$ is a $c$-number function, and $\phi (x)$ acts like a dynamical effective scalar field with anomalous
dimension $2\gamma$. Therefore we can write
\br
\Sigma (p,k) &\sim& \phi (k) \left( \frac{ \Lambda^2 }{p^2}\right)^{\alpha}[1 + b g^2 \ln\left(p^2/ \Lambda^2 \right)]^{-\gamma\cos (\alpha \pi)}
\nonumber \\
&\equiv& \phi (k) \tilde{\Sigma} (p^2)  \, .
\label{eq18}
\er	
As seen in Eq.(\ref{eq18}), working in the true vacuum generates a nontrivial dependence on
the momentum $k$ for our variational parameter $\phi$. The kinetic term for our effective theory
is obtained inserting $\phi (k)$ in the effective action and expanding around $k=0$. The diagrams
contributing to the kinetic part of the energy density are shown in Fig.(\ref{fig2})

\section{The Ginzburg-Landau Lagrangian}

\noindent In order to determine the effective Lagrangian we start computing
the kinetic term contribution, which is given by the polarization
diagrams ($\Pi (k^2,\phi)$) of Fig.(\ref{fig2}). This contribution
is important in our calculation because it will give the correct normalization of the effective fields,
as discussed after Eq.(\ref{eq2}). The renormalization constant for
the scalar composite field is obtained from~\cite{cs},
\be
Z \approx 2 \left.\frac{d\Pi (k^2,\phi ) }{dk^2}\right|_{k^2 =0}  \, .
\label{eq19}
\ee	
The Taylor expansion of $Z$ around $k^2=0$ gives
\be
Z \approx \left. \frac{k^2}{8} g_{\gamma\delta} \frac{\partial}{\partial k_\gamma}\frac{\partial}{\partial k_\delta}\Pi (k^2,\phi )\right|_{k^2 \approx 0}  \, ,
\label{eq20}
\ee	
which after some algebra can be written as
\be 
(Z^{(\alpha)})^{-1} \approx \frac{N_{TC}n_F}{4\pi^2}\int dp^2\frac{(p^2)^2\tilde{\Sigma}^2(p^2)}{(p^2 + \Lambda^2_{TC})^3} \, ,
\label{eq201}
\ee 
where the index $\alpha$ is related to the ansatz of Eq.(\ref{eq18}). 

Using Eq.(\ref{eq18}), considering that the fermions in the loop
have technicolor and flavor numbers equal to $N_{TC}$ and $n_F$, respectively, and after some calculation we obtain
\be
Z^{(0)}  \approx \frac{4 \pi^2 \beta (2\gamma -1)}{N_{TC}n_F}\left[ 1+\frac{\alpha}{\beta (\gamma -1)}+ \, ...\right] \, ,
\label{eq21}
\ee
where $Z^{(0)} $ is the normalization constant obtained performing the kinetic
loop calculation and expanding the result in the limit $\alpha \rightarrow 0$. In Eq.(\ref{eq21})
$\beta = bg^2$. The limit $\alpha \rightarrow 0$
will correspond to the extreme walking limit of our effective Lagrangian. We do
the same calculation for the case $\alpha \rightarrow 1$, obtaining 
\be
Z^{(1)} \approx \frac{8\pi^2}{N_{TC}n_{F}}\left[1 - \frac{\beta\gamma}{\alpha} + ...\right] \, .
\label{eq22}
\ee

Our effective Lagrangian will be given by
\be 
\Omega^{(\alpha )} = \int d^4x\left[\frac{1}{2Z^{(\alpha )}(\phi)} \partial_{\mu}\phi\partial^{\mu}\phi\right]  - \Omega_V^{(\alpha )}  \, ,
\label{eq221}
\ee
where $\Omega_V^{(\alpha)}$ can be written in powers of $\phi$ leading to
\be
\Omega^{(\alpha )} = \int d^4x\left[\frac{1}{2Z^{(\alpha )}(\phi)} \partial_{\mu}\phi\partial^{\mu}\phi   - \frac{\lambda_{4V}^{(\alpha)}}{4}\phi^4 - \frac{\lambda_{6V}^{(\alpha)}}{6}\phi^6 - ...\right]\,,
\label{eq222}
\ee
that after renormalization by $Z^{(\alpha )}$ translates to
\be
\Omega^{(\alpha )}_{R} = \int d^4x\left[\frac{1}{2} \partial_{\mu}\Phi\partial^{\mu}\Phi   - \frac{\lambda_{4VR}^{(\alpha)}}{4}\Phi^4 - \frac{\lambda_{6VR}^{(\alpha)}}{6}\Phi^6 - ...\right]\,.
\label{eq2222}
\ee
In this expression we have defined the renormalized field $\Phi\equiv [Z^{(\alpha )}]^{-\frac{1}{2}}\phi$, and the renormalized couplings for the two limits, $\alpha\rightarrow 0$ and $\alpha\rightarrow 1$, are given, respectively,  by
\begin{widetext}
\br 
&&\lambda^{(0)}_{4VR} \equiv \lambda_{4V}^{(0)} [Z^{(0)}]^2 = \frac{N_{TC}n_{F}}{4\pi^2}[Z^{(0)}]^2
\times  \left[\left(\frac{1}{\beta(4\delta - 1)} +\frac{1}{2}\right)  - \frac{4\alpha}{\beta(4\delta - 1)}\left(\frac{1}{(4\delta - 2)} +2\delta\right)\right]\, ,  \nonumber \\
&& \lambda^{(0)}_{6VR} \equiv \lambda_{6V}^{(0)}[Z^{(0)}]^3 = - \frac{N_{TC}n_{F}}{4\pi^2}\frac{[Z^{(0)}]^3}{\Lambda^2{{}_{TC}}}  \, , 
\er
and
\br 
&&\lambda^{(1)}_{4VR} \equiv \lambda_{4V}^{(1)} [Z^{(1)}]^2 = \frac{N_{TC}n_{F}}{4\pi^2}[Z^{(1)}]^2 
\times  \left[\frac{1}{4}\left(1  + \frac{c\alpha_{{}_{TC}}}{2\pi}\right)  - \frac{\beta}{4\alpha}\left(\delta + \frac{c\alpha_{{}_{TC}}}{8\pi}(4\delta + 1)\right)\right]\, ,  \nonumber \\
&& \lambda^{(1)}_{6VR} \equiv \lambda_{6V}^{(1)} [Z^{(1)}]^3= - \frac{N_{TC}n_{F}}{4\pi^2}\frac{[Z^{(1)}]^3}{7\Lambda^2{{}_{TC}}} \, ,
\label{eq224}
\er
\end{widetext}
where the $\lambda_{nV}^{(\alpha)}$ are the couplings computed in Appendix A.

Note also that besides the absence of a $\phi^2$ term, due to the fact that we assumed that our ansatz satisfies
the linear fermionic self-energy equation, we do not have odd powers of the effective 
$\phi$ field in the potential because we are assuming massless technifermions~\cite{bard}. 

Up to now we have discussed the contributions to the effective Lagrangian that are originated
from the new fermions responsible for the composite scalar state. In models based on the technicolor
idea, the composite scalar boson is made of these new fermions only.
Of course there are models like topcolor~\cite{top} where the top quark has a strong interaction
such as it could supply the scalar composite necessary to the dynamical symmetry breaking
of the electroweak theory. This last possibility would change the contributions that we should
consider to the potential, but as there is no observed signal in the top quark physics up
to now indicating such a possibility, we do not follow this path and consider that our
composite state is formed only by new fermionic degrees of freedom. However, even in this
case we still have other contributions to the effective Lagrangian. The contributions that
we are referring to are the ones coming from ordinary massive quarks and leptons that couple to the
scalar boson. These contributions will be dominated by the heaviest fermion 
(the top quark) and will
generate terms of order $\phi^3$, $\phi^4$, and higher as will be discussed in the sequence.  

\begin{figure}
\centering
\includegraphics[width=0.6\columnwidth]{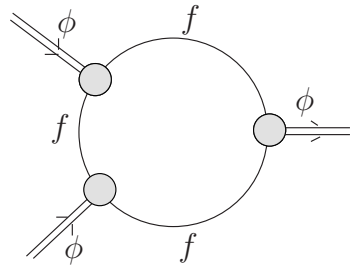}
\caption[dummy0]{Heavy ordinary fermions ($f$) contribution to the trilinear composite ($\phi$) Higgs boson coupling.
The gray blobs are proportional to the effective $ff\phi$ coupling.} 
\label{lamb3}
\end{figure}

\begin{figure}
\centering
\includegraphics[width=0.6\columnwidth]{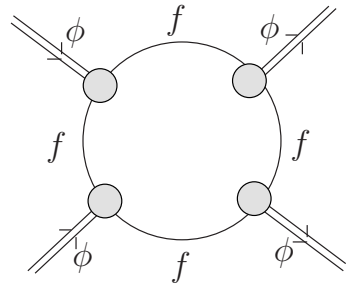}
\caption[dummy0]{Heavy ordinary fermions ($f$) contribution to the quadrilinear composite ($\phi$) Higgs boson coupling.
The vertices are proportional to the effective $ff\phi$ coupling.} 
\label{lamb4}
\end{figure}

The $\phi^3$ and $\phi^4$ contributions to the effective Lagrangian due to the
ordinary massive fermions are given, respectively,
by the diagrams of Figs. \ref{lamb3} and \ref{lamb4}, where the effective $ff\phi$ coupling
is determined through Ward identities as discussed in Refs.~\cite{soni,soni2,doff3}, and it
is easy to verify that such a coupling will be given by
\be
\imath \lambda_{\phi ff} \propto -\imath \frac{g_W \Sigma_f (k)}{2M_W} \, .
\label{eq28}
\ee

Notice that the ordinary fermions masses in composite Higgs models come from a new type of interaction,
that in the most common approach is called extended technicolor interaction (ETC). As we do not know
the dynamics (or model) for this specific interaction, we cannot formally derive their contribution to the effective
action. However, we can compute the effect of ordinary fermions to the effective potential 
as a function of their masses, exactly as performed by Carpenter {\it et al.}~\cite{ressoni}. These contributions
are expected to be small, since the ordinary fermion masses are smaller than the characteristic composite scale ($\Lambda_{TC}$).
The calculation of the $\phi^3$ and $\phi^4$ terms are
presented in Appendix B, where we determine the effective trilinear and quadrilinear couplings (the contributions to $\Omega_V$
are obtained multiplying these couplings by the normalized fields). The couplings are equal to:

a) Trilinear coupling when $\alpha \sim 0 $             
\be
\lambda^{(0)}_{3f} \approx \frac{9g^3_{W}}{32\pi^2}\frac{m_{t}}{\beta(4\delta - 1)}\left(\frac{m_{t}}{M_{W}}\right)^3\left[1 - \frac{4\alpha}{\beta(4\delta - 2)}+ ...\right] \, ,  
\label{eq29}
\ee

b) Trilinear coupling when $\alpha \sim 1 $
\be
\lambda^{(1)}_{3f} \approx \frac{9g^3_{W}}{32\pi^2}\frac{m_{u}}{4}\left(\frac{m_{u}}{M_{W}}\right)^3\left[1 - \frac{\beta(4\delta - 1)}{4\alpha}+ ...\right]\, ,
\label{eq30}
\ee

c) Quadrilinear coupling when $\alpha = 0 $  
\be 
\lambda^{(0)}_{4f} \approx \frac{3g^4_{W}}{64\pi^2 M^4_{W}}\frac{m^4_{t}}{\beta(4\delta - 1)}\, ,
\label{eq31}
\ee    
This result is the same as the one obtained by Carpenter {\it et al.}~\cite{ressoni}.

d) Quadrilinear coupling when $\alpha = 1 $  
\be 
\lambda^{(1)}_{4f} \approx \frac{3g^4_{W}}{64\pi^2 M^4_{W}}\frac{m^4_{u}}{4}\, .
\label{eq32} 
\ee 

The fact that when $\alpha \sim 1$ we introduced the mass $m_u$, as discussed in Appendix B, is an approximation, because in
this case we can only generate light fermion
masses, in order to be consistent with the absence of flavor changing neutral currents. Actually we should say that this last
case is not important and should not be considered, since a relevant contribution would come from heavy fermions, and as far as it is known up to now~\cite{doff2}, such heavy mass could only be naturally 
generated in extreme walking gauge theories.

Observe that the above Eqs.~(\ref{eq29})-(\ref{eq32}) can easily be rewritten in terms of the TC scale $\Lambda_{TC}$ if we use the relation between $M_W$ and the technipion constant $F_\Pi$,
\be
M_W^2=\frac{g_w^2 n_d F_\Pi^2}{4}\,,
\label{eq33}
\ee
where $n_d$ is the number of technifermion doublets and $F_\Pi$ is obtained from the Pagels and Stokar relation~\cite{pagels},  
\be 
F^2_{\Pi} = \frac{N_{{}_{TC}}}{8\pi^2}\int\!\!\frac{dp^2p^2}{(p^2 + \Sigma^2(p^2))^2}\!\!\left[\Sigma^2(p^2) - \frac{p^2}{2}\frac{d\Sigma(p^2)}{dp^2}\Sigma(p^2)\right]\,.
\label{eq34}
\ee 
After transforming the above momentum integral in Eq.~(\ref{eq34}) through a Mellin transformation,
\be
[1+\beta \ln{\frac{p^2}{\Lambda^2_{TC}}}]^{-2\delta}=\frac{1}{\Gamma (2\delta)}\int_0^\infty\,dz\,z^{2\delta -1}e^{-z}(\frac{p^2}{\Lambda^2_{TC}})^{-\beta z}\, .
\ee
and using  Eq.~(\ref{eq12}) and the following expression for the factor $Z$
\be 
(Z^{(\alpha)})^{-1} \approx \frac{N_{TC}n_{F}}{4\pi^2}\frac{1}{\Gamma(2\delta)}\int^{\infty}_{0}dz\frac{z^{2\delta - 1}e^{-z}}{(2\alpha + \beta z)}\, ,
\ee 
we can rewrite the equation for $F_\Pi$ in terms of $Z^{(\alpha)}$, which leads to
\be 
n_d F^2_{\Pi} = \left( 1 + \frac{\alpha}{2}\right)\frac{\Lambda^2_{TC}}{Z^{(\alpha)}}\, .
\label{eq35}
\ee 
Equation(\ref{eq35}) is an interesting example of how the technipion decay constant varies with the theory dynamics (or
with $\alpha$). Notice that if we change the dynamics of the theory we cannot obtain $F^2_{\Pi}$ just with a  simple scaled QCD. 
Another interesting fact is also the relation between $F^2_{\Pi}$ and $Z^{(\alpha)}$. 
Since the fields in the effective Lagrangian are normalized by different
powers of $Z^{(\alpha)}$ (or powers of $F^2_{\Pi}$, that also varies with $\alpha$), and since $F^2_{\Pi}$ is fixed by
the weak gauge boson masses, we verified that the behavior of the effective theory is quite different according to
the different limits of the $\alpha$ parameter.

\section{Results} 

The full effective Lagrangian for the composite Higgs system will be given by 
\be 
\Omega = \int d^4x\left[\frac{1}{2Z^{(\alpha)}} \partial_{\mu}\phi\partial^{\mu}\phi \right] - \Omega_V  \, . 
\label{eq36}
\ee
Introducing the normalized field 
\be 
\Phi = \phi [Z^{(\alpha)}]^{-1/2}  \, ,
\ee 
we can write
\br 
&&\Omega^{(\alpha)}_R = \int d^4x \left[\frac{1}{2} \partial_{\mu}\Phi\partial^{\mu}\Phi  - \frac{\lambda^{(\alpha)}_{3fR}}{3}\Phi^3 \right. \nonumber \\&&- \left. \frac{({\lambda^{(\alpha)}}_{4VR}+{\lambda^{(\alpha)}}_{4fR})}{4}\Phi^4 
- \frac{{\lambda^{(\alpha)}}_{6VR}}{6}\Phi^6 + ... \right]
\label{eq37}
\er

The coupling constants that appear in Eq.(\ref{eq37}) are the ones obtained in the previous section. It must be noticed
that the couplings originated from the ordinary fermion masses are smaller than the ones generated from the techniquarks
effective potential. For example,
\be
\frac{\lambda_{4V}^{(0)}}{\lambda^{(0)}_{4f}} \approx \frac{(N_{TC}n_{F})^3}{12(16\pi^2\beta(2\gamma-1))^2}\left(\frac{\Lambda_{{}_{TC}}}{m_{t}}\right)^4\,.
\label{eq38}
\ee
The $\beta(2\gamma-1)$ factor appearing in the denominator is usually of ${\cal O}(1)$ for several gauge groups. Assuming a $SU(4)_{{}_{TC}}$
technicolor theory with $n_F = 14$~\cite{comment}, in the case when $\alpha \rightarrow 0$, we will obtain a ratio of the following order 
\be
\frac{\lambda_{4V}^{(0)}}{\lambda^{(0)}_{4f}} \approx {\cal O}(10) \, .
\label{eq39}
\ee
In the case when $\alpha \rightarrow 1$ the difference can be even larger.
This means that we can neglect the ordinary massive fermions contribution to $\Omega_R^{(\alpha)}$ 
(proportional to $\lambda^{(\alpha)}_{nf}$) compared to
the one of techniquarks (proportional to $\lambda^{(\alpha)}_{nV}$). The only exception is the 
$(\lambda^{(\alpha)}_{3fR}/3)\Phi^3$ term, which is small but is the leading term of this order in the 
effective action and introduces some effect in the scalar mass calculation.

We can now compute the scalar mass which is determined from the following equation: 
\be 
M^{2(\alpha)}_{{}_{\Phi}} = \frac{\partial^2\Omega^{(\alpha)}_R}{\partial\Phi^2}|_{{}_{{}_{\Phi=\Phi_{min}}}} \, .
\label{eq40}
\ee 

After neglecting terms proportional to $\lambda_6$ and of higher order when substituting the minimum value
in the potential we obtain 
\be 
M^{2(\alpha)}_{{}_{\Phi}} \approx 2\lambda^{(\alpha)}_{4VR}\left(\frac{\lambda^{(\alpha)}_{4VR}}{\lambda^{(\alpha)}_{6VR}}\right) + 5\lambda^{(\alpha)}_{3fR}\left(\frac{\lambda^{(\alpha)}_{4VR}}{\lambda^{(\alpha)}_{6VR}}\right)^{1/2}.
\label{eq41}
\ee 

\begin{figure}
\centering
\includegraphics[width=0.99\columnwidth]{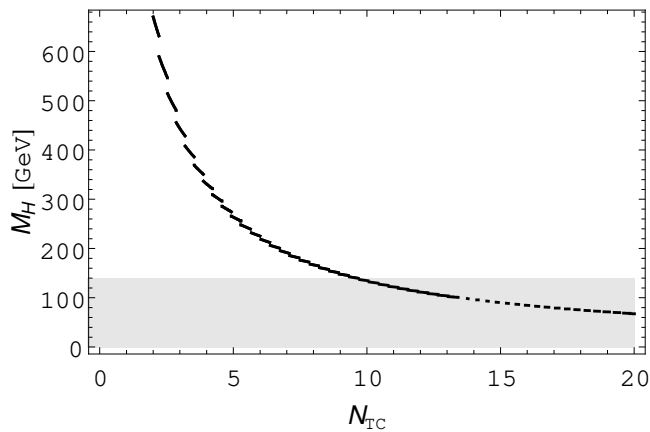}
\caption[dummy0]{Higgs mass as a function of $N_{TC}$ in the extreme walking technicolor regime. The shaded band is the experimentally excluded region~\cite{cdf}.} 
\label{lamb7}
\end{figure}

With Eq.(\ref{eq41}) we can compute numerically the Higgs boson mass in the extreme walking behavior ($\alpha \rightarrow 0$) and the
result is plotted in Fig.(\ref{lamb7}). Notice that as we go to larger values of $N_{TC}$ while keeping a slowly TC running coupling
constant (a $\beta_{TC}$ function close to zero) we verify that the current experimental limit on the Higgs boson mass does not allow
us to have a technicolor gauge group arbitrarily large ($N_{TC}< 10$). The possibility that a composite Higgs boson can be as light as the
present experimental limit has been already noticed in a series of papers~\cite{LCH}.
The authors of these papers particularly discuss a more interesting 
case where the walking behavior is obtained in theories where the 
fermions are in higher dimensional representations of the technicolor group, turning unnecessary the introduction of a quite large number of fermions, as
happens in the case where the fermions are in the fundamental representation. Moreover,  it was also shown that exactly for the extreme walking case these theories, with a light composite Higgs, are totally in agreement with the precision electroweak measurements~\cite{LCH}. 
In obtaining Fig.(\ref{lamb7}) we have used the
$\beta$ function up to two loops, where $n_F$  for each $SU(N_{TC})$  has
to  be fixed accordingly, i.e., $n_F=8,\,11,\,14,...$ for $N_{TC}=2,\,3,\,4,...$

Considering the smallest possible non-Abelian unitary technicolor gauge group, i.e. $SU(2)_{TC}$, we can observe from Fig.(\ref{lamb7})
that, in the extreme walking regime, the Higgs boson mass has an upper limit of about ${\cal O}(700)$ GeV. In order to have
models with dynamical symmetry breaking along the technicolor idea without the problems of neutral flavor changing currents,
the walking scenario seems to be the most promising possibility~\cite{walk}. In this context our result implies a crucial test for the walking
technicolor hypothesis since such mass values may be promptly assessed at LHC.    

Let us consider the limit $\lambda^{(\alpha)}_{3fR} \rightarrow 0$. In this case we obtain the conventional result for $M^{2(\alpha)}_{{}_{\Phi}}$
given by the effective potential 
\be 
M^{2(\alpha)}_{{}_{\Phi}} \approx 2\frac{[\lambda^{(\alpha)}_{4V}]^2}{\lambda^{(\alpha)}_{6V}} \, .
\label{eq42}
\ee

\begin{figure}
\centering
\includegraphics[width=0.99\columnwidth]{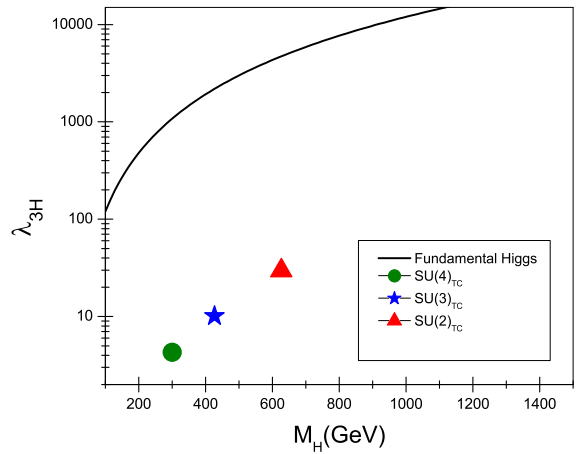}
\caption[dummy0]{Trilinear scalar coupling as a function of composite boson mass. We plot a solid line with the standard model
value for this relation and show the expected values for a composite Higgs boson based on $SU(2)_{TC}$, $SU(3)_{TC}$, and $SU(4)_{TC}$ models, with
$n_f = 8, \,\, 11, \,\, 14$, respectively.} 
\label{lamb6}
\end{figure}

We can observe that the top quark mass ($m_{t} \sim 175 GeV$) will usually give a contribution of the order of $10\%$ of the composite
Higgs boson mass through the trilinear Higgs boson coupling. We show in Fig.(\ref{lamb6}) some values for the Higgs mass versus the
trilinear coupling for some technicolor models already discussed in the literature~\cite{walk}. The points that we have chosen in Fig.(\ref{lamb6})
correspond to extreme walking technicolor theories, and we expect the possible range of couplings and masses for other
$\alpha$ values to be located between these points and the standard model curve~\cite{doff2}.

Finally, in the limit $\alpha \rightarrow 1$ we simply obtain scalar boson masses in the TeV region as usual, but these models
are known to be plagued by unwanted flavor changing neutral currents.

\section{Conclusions}

We have computed an effective action for a composite Higgs boson system formed
by new fermions belonging to a general technicolor non-Abelian gauge theory. 
The calculation is based on an effective action for composite operators. The novelty
is that the effective action is computed with the help of a quite general self-energy that 
depends on a certain parameter ($\alpha$), which, when variated from $0$ to $1$, provides an
interpolation of the fermionic self-energy from the extreme walking technicolor behavior up
to the self-energy expression that obeys the standard operator product expansion. This means
that our calculation is quite general in the sense that choosing values for $\alpha$,
which is equivalent to choosing different dynamics for the strong interaction forming the composite scalar boson,
we can obtain the different mass and couplings of the effective theory.

There are two other improvements in our calculation. The first one is the calculation of the
kinetic term of the effective theory. This term appears with a coefficient that differs from the
standard parametrization of the kinetic term of a scalar Lagrangian. When the effective Lagrangian
is normalized to reproduce a standard scalar effective field theory we also must modify the
remaining terms, leading to a nontrivial change of the scalar self-couplings. The second improvement
is that we also consider the effect of ordinary massive fermions to the effective Lagrangian. This
contribution is usually neglected, and is indeed small except by the contribution of the heavy 
top quark. An ordinary massive fermion contribution is also important because it introduces
odd powers of scalar field self-couplings in the theory, i.e., the trilinear composite scalar 
self-coupling is originated from the loop of the top quark, while the quadrilinear self-coupling is 
dominated by the techniquarks interactions as well as, in a minor extent, from the top quark 
loop contribution. Of course, this result would change if there is a fourth ordinary
fermion family or if the techniquarks have a large current mass (above the TC scale).

With the general fermionic self-energy (or composite state wave function) we computed
the effective Lagrangian presenting the results for $\alpha =0$ and $\alpha =1$, which correspond
to the limits of the extreme walking technicolor theory and the standard view of the technicolor
theory that can be obtained by scaled QCD. For other $\alpha$ values the scalar mass and self-couplings 
are located between the ones obtained for the extreme cases ($0$ and $1$). In the case
of an extreme walking behavior ($\alpha \rightarrow 0$), we
obtain an upper limit for the Higgs boson mass, ($M_H \leq {\cal O} (700)$~GeV for $SU(2)_{TC}$), and
the experimental data on the Higgs boson mass constrain $SU(N)_{TC}$ technicolor gauge groups to be smaller 
than $SU(10)_{TC}$, whereas when $\alpha \rightarrow 1$ the scalar
mass is expected to be much heavier. Therefore we agree with the earlier results of Ref.~\cite{LCH} that
we may have quite a light composite Higgs scalar boson in the case of extreme
walking TC theories.

It is fair to mention that another source of
uncertainty in our approach, besides that assumed in the SDE, is that we are showing results for the extreme walking behavior ($\alpha \rightarrow 0$), for which we consider the
$\beta$ function up to 2 loops. This obviously constrains the number of fermions introduced in our computation of the effective potential. Higher loops certainly change the number of fermions needed to get the walking behavior, implying a change in our numerical results for the triple and quartic gauge couplings as well as the Higgs mass. It is possible that going to further orders of the beta function could modify the specific shape of the Higgs mass function shown in Fig.5 and shift the couplings relative to the expected results of the standard model plotted in Fig.6. Given the degree of approximations we have already assumed in computing the effective potential, we thought it was reasonable to truncate the beta function to the order that its coefficients are universal. However, it is interesting to notice that a complete all orders beta function obtained in Ref.~\cite{RS} could be used in a more general approach and also when different representations are considered and could be helpful in developing an extended analysis in a future work.

In theories where the scalar Higgs boson is composite we need new ``extended technicolor" interactions
in order to give masses to the ordinary fermions. As far as we know there is no phenomenologically 
viable ETC model and its effect enters in our effective Lagrangian parametrized in the massive
ordinary fermion contributions. This contribution is important, as discussed above, because it
is responsible for the trilinear scalar coupling and we expect that other ETC contributions
decouple from the effective Lagrangian. The ordinary fermion contributions to the effective Lagrangian
are roughly 1 order of magnitude smaller than the one of techniquarks. This is an expected behavior
since their masses are smaller than the TC mass scale.   
 
At present the walking technicolor models seem to be the most promising possibility for dynamically broken gauge
theories. Therefore if this scenario is appropriate to describe the dynamics of symmetry breaking (with a unitary
gauge group in the fundamental representation), our limit implies 
that the scalar composite boson should be observed at LHC with a mass up to $700$~GeV, a quite interesting outcome considering a composite nature for the Higgs boson.
 
\acknowledgements
This research was partially supported by the Conselho Nacional de Desenvolvimento
Cient\'{\i}fico e Tecnol\'ogico (CNPq).

\appendix
\section{$\phi^4$ and $\phi^6$ contributions to $\Omega_V$}
In this appendix we compute the $\phi^4$ and $\phi^6$ terms of the effective
potential $\Omega_V$. We start from the effective action up to two loops (see Eq.(\ref{eq11})):
\br
\Omega_V^{(\alpha)} &=& i Tr \ln (1- \Sigma S_0) + \frac{1}{2} i Tr \Sigma S_0 \Sigma S_0 \nonumber \\
&+&  \frac{1}{2} i Tr S_0\Sigma S_0 \Sigma S_0\Gamma S_0 \Sigma S_0 \Sigma S_0 \Gamma D_0 \, , 
\label{eqA1}
\er
where the last term comes from the two-loop contribution.

Expanding the term proportional to $\ln(1- \Sigma S_0)$, considering the propagators, vertices and the
$\Sigma$ ansatz with the momentum dependence of Eq.(\ref{eq18}), we obtain 
\begin{widetext}
\br
&&\Omega_V^{(\alpha)} = \frac{N_{TC}n_{F}}{16\pi^2}\frac{1}{\Gamma(4\delta)}\int \frac{dz z^{4\delta - 1}e^{-z}}{(4\alpha + \beta z)}Tr(\phi^4) +
\frac{N_{TC}n_{F}}{16\pi^2}\frac{3\alpha_{{}_{TC}}c}{4\pi(2 + 4\alpha)}\frac{4}{\Gamma(4\delta  + 1)}\int \frac{dz z^{(4\delta+1) - 1}e^{-z}}{(4\alpha + \beta z)}Tr(\phi^4) \nonumber \\
&&+
\frac{N_{TC}n_{F}}{16\pi^2}Tr\left[\phi^4\left(\sum_{{}_{{}_{m=1}}}\frac{\left(\frac{\phi^2}{\Lambda^2{{}_{TC}}}\right)^{m}}{(2m + 4)}\frac{(-1)^{m}}{(m + (2m + 4)\alpha)}\right)\right] \, ,
\label{eqA2}
\er
\end{widetext}
where the contribution of ${\cal{O}}(\Sigma^2)$ is canceled between the two first terms of Eq.(\ref{eqA1}),
and the last term is what remains of the $\ln(1- \Sigma S_0)$ expansion after cancellation of the $\Sigma^2$
contribution and extraction of the $\Sigma^4$ contribution. 

We can compute Eq.(\ref{eqA2}) in the limits $\alpha = 0$ and $\alpha=1$.  
In the case $\alpha \approx 0$ we have 
\begin{widetext}
\br 
\Omega_V^{(0)} &=& 
\frac{N_{TC}n_{F}}{16\pi^2}\left[\left(\frac{1}{\beta(4\delta - 1)} + \frac{1}{2}\right)  -\frac{4\alpha}{\beta(4\delta - 1)}\left(\frac{1}{(4\delta - 2)} + 2\delta\right)\right]Tr(\phi^4) 
\nonumber \\ 
&&+ \frac{N_{TC}n_{F}}{16\pi^2}\left[-\Lambda^4_{{}_{TC}}Tr\left(\frac{\phi^2}{\Lambda^2_{{}_{TC}}}\right)  +\frac{\Lambda^4_{{}_{TC}}}{2}Tr\left(\frac{\phi^4}{\Lambda^4_{{}_{TC}}}\right)\right] 
+ \frac{N_{TC}n_{F}}{16\pi^2}\left[Tr\left[\left(1 - \frac{\phi^4}{\Lambda^4_{{}_{TC}}}\right)\ln\left(1 +\frac{\phi^2}{\Lambda^2_{{}_{TC}}}\right)\right]\right].
\label{eqA3}
\er
If we assume $\frac{\phi^2}{\Lambda^2_{{}_{TC}}} << 1$ we obtain
\br
\Omega^{(0)}_V = \!\!\!\!\!\!&&\frac{N_{TC}n_{F}}{16\pi^2}\left[\left(\frac{1}{\beta(4\delta - 1)} + \frac{1}{2}\right)  - \frac{4\alpha}{\beta(4\delta - 1)}\left(\frac{1}{(4\delta - 2)} + 2\delta\right)\right]Tr(\phi^4)
-\frac{N_{TC}n_F}{16\pi^2}\left[\frac{2}{3\Lambda^2_{{}_{TC}}}\right]Tr(\phi^6) + O(Tr(\phi^8)) ... \nonumber \\
\label{eqA4}
\er 
In the limit $\alpha \approx 1$ we have 
\br 
&&\Omega_V^{(1)} = 
\frac{N_{TC}n_{F}}{16\pi^2}\left[\frac{1}{4}\left(1  + \frac{c\alpha_{{}_{TC}}}{2\pi}\right)  - \frac{\beta}{4\alpha}\left(\delta + \frac{c\alpha_{{}_{TC}}}{8\pi}(4\delta + 1)\right)\right]Tr(\phi^4) 
\nonumber \\ && +
\frac{N_{TC}n_{F}}{16\pi^2}\left[ \Lambda^4_{{}_{TC}}Tr\left(\frac{\phi^2}{\Lambda^2_{{}_{TC}}}\left[2 - 3{}_{2}F_{1}\left(1,\frac{1}{3};\frac{4}{3}; -\frac{\phi^2}{\Lambda^2_{{}_{TC}}}\right)\right]\right)\right]  + 
\frac{N_{TC}n_{F}}{16\pi^2}\left[ - \frac{\Lambda^4_{{}_{TC}}}{4}Tr\left(\frac{\phi^4}{\Lambda^4_{{}_{TC}}}\right) + \Lambda^4_{{}_{TC}}\ln\left(1 + \frac{\phi^2}{\Lambda^2_{{}_{TC}}}\right)\right]\,,\nonumber \\
\label{eqA5} 
\er
where ${}_pF_{q}(a_1,...,a_p;b_1,...b_q;x)$ is the hypergeometric function.

\par Again assuming $\frac{\phi^2}{\Lambda^2_{{}_{TC}}} << 1$ we obtain the following $\phi^4$ and $\phi^6$ contributions to $\Omega_V^{(1)}$   
\br 
\Omega_V^{(1)} = \!\!\!\!\!\!&&\frac{N_{TC}n_F}{16\pi^2}\left[\frac{1}{4}\left(1  + \frac{c\alpha_{{}_{TC}}}{2\pi}\right)  - \frac{\beta}{4\alpha}\left(\delta + \frac{c\alpha_{{}_{TC}}}{8\pi}(4\delta + 1)\right)\right]Tr(\phi^4)
-\frac{N_{TC}n_F}{16\pi^2}\left[\frac{2}{21\Lambda^2_{{}_{TC}}}\right]Tr(\phi^6) + O(Tr(\phi^8)) ...  \nonumber \\
\label{eqA6}
\er
\end{widetext}
From Eqs.(\ref{eqA4}) and (\ref{eqA6}) we can read the values of the couplings $\lambda^{(0)}_{4V}$ and 
$\lambda^{(1)}_{4V}$, which are given, respectively, by
\be 
\lambda_{4V}^{(0)} \approx \frac{N_{TC}n_{F}}{16\pi^2}\left(\frac{1}{\beta(4\delta - 1)} + \frac{1}{2}\right) \, ,
\label{eqA7}
\ee
\be
\lambda_{4V}^{(1)} \approx \frac{N_{TC}n_{F}}{16\pi^2}\frac{1}{4}\left(1  + \frac{c\alpha_{{}_{TC}}}{2\pi}\right).
\label{eqA8}
\ee 
The $(1/2)$ factor at the end of Eq.(\ref{eqA7}) comes from the two-loop contribution.

In the same way, as done above for the $\lambda^{(\alpha)}_{4V}$ coupling, we can easily obtain the
$\lambda^{(\alpha)}_{6V}$ from Eqs.(\ref{eqA4}) and (\ref{eqA6}). These results are the ones shown in
the Sec. III.

\section{Trilinear and quadrilinear couplings originated from ordinary fermions}

The trilinear and quadrilinear couplings that are originated from the ordinary massive fermions are obtained
from the calculation of Fig.(\ref{lamb3}) and Fig.(\ref{lamb4}), respectively. 
Assuming that the coupling $\phi \bar{f}f$, of the composite Higgs scalar boson $\phi$ to the ordinary fermions, 
at large momentum $p^2$  is given by~\cite{soni2}
$$
\lambda_{\phi ff} \approx - \frac{g_{W}\Sigma_{f}(p^2)}{2M_{W}}\,,
$$
\noindent we obtain 
\be 
\lambda^{(\alpha)}_{4f} \approx \frac{1}{64\pi^2}\frac{g^4_{W}n_FN_{c}}{(M_{W})^4}\int\frac{dp^2 p^6 \Sigma^4_{f}(p^2)}{(p^2 + m^2_{f})^4}, 
\ee 
\noindent  where in this  expression $\Sigma_{f}(p^2)$ is parametrized by the ansatz of Eq.(\ref{eq12}). Moreover,  the infrared cut off $\Lambda$,  which is the characteristic scale of the mass generation in Eq.(\ref{eq12}), in this case will be identified with $\Lambda = m_{f}$ exactly
as performed in Ref.~\cite{soni2}. After some calculation  we can write, in the limit $\alpha = 0$, the following quadrilinear coupling 
\be 
\lambda^{(0)}_{4f} \approx \frac{3g^4_{W}}{64\pi^2 M^4_{W}}\frac{m^4_{f}}{\beta(4\delta - 1)}.
\label{eqB1}
\ee    
\noindent The largest contribution comes from the heaviest fermion, which can be identified with the top quark $(m_{f} = m_{f}(0) \approx m_{t})$ or the lepton tau $(m_{f} = m_{f}(0) \approx m_{\tau})$, if we consider leptons. We do the same calculation for the case $\alpha = 1$, obtaining
\be 
\lambda^{(1)}_{4f} \approx \frac{3g^4_{W}}{64\pi^2 M^4_{W}}\frac{m^4_{f}}{4} \, .
\label{eqB2} 
\ee 
\noindent The self-energy solution, in this specific limit, cannot generate large fermion masses~\cite{doff2} (without generating large flavor
changing neutral currents). Therefore we can expect that $(m_{f} = m_{f}(1)\approx m_{u})$ or $(m_{f} = m_{f}(1) \approx m_{e})$. The trilinear self-coupling of the composite Higgs bosons with the ordinary fermions can be obtained in the same  way, and the result is
\be 
\lambda^{(\alpha)}_{3f} \approx \frac{3g^3_{W}n_FN_{c}}{(M_{W})^3}\frac{1}{32\pi^2}\int\frac{dp^2 p^4 \Sigma^4_{f}(p^2)}{(p^2 + m^2_{f})^3}.
\label{eqB3}
\ee
\noindent where for $\alpha \sim 0 $ and  $\alpha \sim 1 $ we obtain 
\be
\lambda^{(0)}_{3f} \approx \frac{9g^3_{W}}{32\pi^2}\frac{m_{f}(0)}{\beta(4\delta - 1)}\left(\frac{m_{f}(0)}{M_{W}}\right)^3\left[1 - \frac{4\alpha}{\beta(4\delta - 2)}+ ...\right]\,,
\label{eqB4}
\ee
\be
\lambda^{(1)}_{3f} \approx \frac{9g^3_{W}}{32\pi^2}\frac{m_{f}(1)}{4}\left(\frac{m_{f}(1)}{M_{W}}\right)^3\left[1 - \frac{\beta(4\delta - 1)}{4\alpha}+ ...\right]\,,
\label{eqB5}
\ee
The couplings shown in Eqs.(\ref{eqB1}) to (\ref{eqB5}) are the ones appearing in Eq.(\ref{eq29}) to (\ref{eq32}).

\end{document}